\documentclass[preprint,bm,amssymb,amsmath,floatfix,showkeys,10pt,tighten]{revtex4}
\usepackage{epsfig}
\usepackage{psfig}
\begin{document}
\title{Charge stripes due to electron correlations in the two-dimensional
spinless Falicov-Kimball model}
\date{\today}
\author{R. Lema\'nski}
\email{lemanski@int.pan.wroc.pl}
\affiliation{Institute of Low Temperature and Structure Research, Polish
Academy of Sciences, Wroc\l aw, Poland}
\author{J.~K.~Freericks}
\email{freericks@physics.georgetown.edu}
\homepage{http://www.physics.georgetown.edu/~jkf}
\affiliation{Department of Physics, Georgetown University, Washington, DC
20057}
\author{G. Banach}
\affiliation{Daresbury Laboratory, Cheshire WA4 4AD, Daresbury, United Kingdom}

\begin{abstract}
We calculate the restricted phase diagram for the Falicov-Kimball model
on a two-dimensional square lattice.  We consider the limit where the
conduction electron density is equal to the localized electron density,
which is the limit related to the $S_z=0$ states of the Hubbard model.
After considering over 20,000 different candidate phases (with a unit
cell of 16 sites or less) and their
thermodynamic mixtures, we find only about 100 stable phases in the
ground-state phase diagram.  We analyze these phases to describe where
stripe phases occur and relate these discoveries to the physics behind
stripe formation in the Hubbard model.
\end{abstract}
\pacs{71.10.-W, 71.30.+h, 71.45.-d, 72.10.-d}
\keywords{charge-stripes--Falicov-Kimball--Hubbard--phase-diagram}
\maketitle

\section{Introduction}

We find it fitting to write a paper on the spinless Falicov-Kimball (FK) 
model~\cite{falicov_kimball_1969}
to celebrate Elliott Lieb's seventieth birthday.  Elliott, and his 
collaborators, provided two seminal results on this model: (i) the first,
with Tom Kennedy, proved that there was a finite temperature phase transition
to a checkerboard charge-density-wave (CDW) phase in two or more dimensions
for the symmetric half-filled case~\cite{kennedy_lieb_1986,lieb_1986}, 
and (ii) the second, with Daniel Ueltschi
and Jim Freericks, proved that the segregation principle holds for all
dimensions~\cite{freericks_lieb_ueltschi_2002a,freericks_lieb_ueltschi_2002b}
(which states that if the total particle density is less than
one, then the ground state is phase separated if the interaction strength 
is large enough~\cite{freericks_falicov_1990}). The Kennedy-Lieb 
result (along with an independent 
Brandt-Schmidt paper~\cite{brandt_schmidt_1986,brandt_schmidt_1987}) 
inspired dozens of follow-up papers by researchers
across the world.  The Freericks-Lieb-Ueltschi paper generalized Lemberger's
proof~\cite{lemberger_1992}
from one dimension to all dimensions, which finally proved
the decade old Freericks-Falicov conjecture~\cite{freericks_falicov_1990}.  
Both papers are important,
because they are the only examples where long-range order and phase
separation can be proved to occur in a correlated electronic system.

The Falicov-Kimball model has an interesting history too.  Leo Falicov and John 
Kimball invented the spin-one-half  version of the model in 1969 to describe
metal-insulator transitions of rare-earth compounds~\cite{falicov_kimball_1969}.
It turns out that John
Hubbard actually ``discovered'' the spinless version of the FK model
four years earlier in 1965~\cite{hubbard_III_1965}, when he developed the 
alloy-analogy solution to
the Hubbard model~\cite{hubbard_I_1963}
(the so-called Hubbard III solution).  This latter version
was rediscovered by Kennedy and Lieb in 1986~\cite{kennedy_lieb_1986}
when they formulated it as
a simple model for how crystallization can be driven by the Pauli exclusion
principle.

In this contribution, we focus on another problem that can be analyzed in
the FK model---the problem of stripe formation in two dimensions.  The question
of the relation between charge stripes, correlated electrons, and 
high-temperature superconductivity has been asked ever since static stripes
were first seen in the nickelate~\cite{nickel1,nickel2,nickel3,nickel4}
and cuprate~\cite{copper1,copper2,copper3,copper4} materials starting in 
1993.  Two schools
of thought emerged to describe the theoretical basis for stripe formation in the
Hubbard model.  The Kivelson-Emery scenario~\cite{kivelson_emery1,%
kivelson_emery2,kivelson_emery3,kivelson_emery4}
says that at large $U$ the Hubbard
model is close to a phase separation instability but the long-range Coulomb
force restricts the phase separation on the nanoscale; a compromise results
in static stripe-like order.  The Scalapino-White 
scenario~\cite{scalapino_white1,scalapino_white2,%
scalapino_white3,scalapino_white4} says that stripes
can form due to a subtle balance between kinetic-energy effects and 
potential-energy effects, mediated by spin fluctuations.  No long range
Coulomb interaction or phase separation is needed to form these stripes.
There are numerous numerical studies that have tried to shed light onto this
problem.  Unfortunately, they have conflicting results.  High temperature
series expansions on the related $t-J$ model~\cite{putikka_luchini_rice_1992,%
putikka_luchini_2000} show that phase separation
exists, but only when $J$ is large enough, so it is not present in the
large-correlation-strength limit of the Hubbard 
model (where $J\rightarrow 0$).  Monte Carlo 
calculations~\cite{qmc1,qmc2,qmc3} and exact 
diagonalization studies~\cite{ed1,ed2}
give different results: some calculations predict
the stripe formation to occur, others show a linkage between the stripe 
formation and the boundary conditions selected for the problem.  
Mean-field-theory analyses~\cite{zaanen_oles_1996,oles_1999}
seem to predict stripe formation without any
phase separation.  One way to make sense of these disparate results is that
both the energy of the intrinsic stripe phases and the energy for phase
separation are quite close to one another, so any small change (induced by 
finite-size effects, statistical errors, effects of correlations not included
in the perturbative expansions, or due
to terms dropped or added to the Hamiltonian) can have a large 
effect on the phase diagram by producing a small relative change
in the energetics of the different many-body states (because of their near
degeneracy).

We take an alternate point of view here.  We choose to examine a model that
can be analyzed rigorously, and can be continuously connected to the Hubbard
model.  We choose the regime that connects directly with the $S_z=0$ states of
the Hubbard model.  The model we analyze is the spinless Falicov-Kimball
model on a square lattice
\begin{equation}
\mathcal{H}=-t\sum_{\langle ij\rangle}(c_i^\dagger c_j+c_j^\dagger c_i)+U
\sum_{i=1}^{|\Lambda|} w_ic_i^\dagger c_i,
\label{eq: ham}
\end{equation}
where $c_i^\dagger$ ($c_i$) creates (destroys) a spinless conduction electron
at site $i$, $t$ is the hopping matrix element ($\langle ij\rangle$ denotes
a summation over nearest-neighbor pairs on a square lattice), $w_i=0$ or 1 is
a classical variable denoting the localized electron number at site $i$, and
$U$ is the on-site Coulomb interaction energy. The Fermionic operators
satisfy anticommutation relations $(c^\dagger_i,c^\dagger_j)_+=0$,
$(c_i,c_j)_+=0$, and $(c^\dagger_i,c_j)_+=\delta_{ij}$.  The symbol $|\Lambda|$
denotes the total number of lattice sites in the square lattice $\Lambda$.
We will always be dealing with periodic configurations of localized electrons,
which means we can always consider our lattice to have a large but finite number
of lattice sites and periodic boundary conditions. A short presentation of these
results has already appeared~\cite{lemanski_freericks_banach_2002}.

The Falicov-Kimball model can be viewed as a Fermionic quantum analogue of
the Ising model, while the Hubbard model can be viewed as the Fermionic
quantum analogue of the Heisenberg model (indeed in the large-$U$ limit at
half filling, the Falicov-Kimball model maps onto an effective nearest-neighbor
Ising model, while the Hubbard model maps onto an effective nearest-neighbor
Heisenberg model).  The way to link the Falicov-Kimball model to the 
Hubbard model is to imagine a generalization of the Hubbard model where the
down-spin hopping matrix element differs from the up-spin hopping matrix
element.  Then as $t_\downarrow\rightarrow 0$, the down spins become heavy
and are localized on the lattice; the quantum-mechanical ground state is
determined by the configuration of down-spin electrons that minimizes the 
energy of the up-spin electrons.  This is precisely the Falicov-Kimball
model!

In order to maintain the connection to the Hubbard model in zero
magnetic field, we must choose the conduction electron density
$\rho_e=\sum_{i=1}^{|\Lambda|}\langle c^\dagger_ic_i\rangle/|\Lambda|$
to be equal to the localized electron density $\rho_f=\sum_{i=1}^{|\Lambda|}w_i/
|\Lambda|$, which we do here.  We study the evolution from the checkerboard 
phase at half filling ($\rho_e=\rho_f=1/2$) to the segregated phase, which 
appears when $\rho_e=\rho_f$ is small enough.  Since these two phases are
drastically different from each other, the transition is likely to include many
different intermediate phases.  Indeed, the ground state phase diagram of
the Falicov-Kimball model can be quite complex. There are many different
periodic phases that can be stabilized for different values of $U$ or
$\rho_e=\rho_f$.  As $U$ becomes large though, the phase diagram simplifies,
as the segregated phase becomes the ground state for wider and wider ranges
of the electron densities.

\section{Formalism}

Our strategy to examine the FK model is a brute-force approach which is
straightforward to describe, but tedious to carry out.  We employ the so-called
restricted phase diagram approach, where we consider the grand-canonical
thermodynamic potential
of the system for all possible periodic phases of the localized
electrons, selected from a finite set of candidate phases.  In this work,
we consider 23,755 phases, which corresponds to the set of all inequivalent
phases with a unit cell that includes 16 or fewer lattice sites.  In order
to calculate the thermodynamic potential, we first must determine 
the electronic 
band structure for the conduction electrons for each candidate periodic phase.
We employ a Brillouin-zone grid of $110\times 110$ momentum points for each
bandstructure.  This requires us to diagonalize up to $16\times 16$ matrices
at each discrete momentum point in the Brillouin zone and results in at most 16
different energy bands. Hence, our calculations can be viewed as finite-size
cluster calculations with cluster sizes ranging from $110\times 110\times 1$ up 
to $110\times 110\times 16$ depending on the number of atoms in the unit cell.
An example of such a bandstructure is shown in 
Fig.~\ref{fig: bands}. The eigenvalues of the 
band structure are summed to determine the ground-state energy for
each number of conduction electrons.  
The Gibbs thermodynamic potential
is then calculated for all possible values of the chemical potentials
of the conduction and localized electrons through the formula
\begin{equation}
G(\{w_i\})=\frac{1}{12100N_0}
\sum_{\epsilon_j<\mu_e} \epsilon_j(\{w_i\})-\mu_e\rho_e -\mu_f\rho_f, 
\label{eq: gibbs}
\end{equation}
with $\mu_e$ and $\mu_f$ denoting the chemical potentials for the conduction
and localized electrons, respectively, and $N_0$ denoting the number of 
atomic sites in the unit cell for the given configuration of localized 
electrons. The symbol $\epsilon_j(\{w_i\})$ denotes the energy eigenvalues of 
the band structure for the given configuration of localized electrons.
Since the thermodynamic potential is concave,
the phase diagram can be directly determined in the chemical potentials 
plane~\cite{watson_lemanski_1995,gajek_jedrzejewski_lemanski_1996a,%
gajek_jedrzejewski_lemanski_1996b}.
Next, we convert the grand canonical ensemble into a canonical ensemble to
determine the ground-state phase diagram as functions of $\rho_e$ and $\rho_f$.
We find the ground state is often a phase separated mixture of two or three
different phases, which can be periodic phases, or the segregated phase.
This step of the analysis is quite complicated, because small
areas of stability in the grand canonical phase diagram can correspond to large
regions in the canonical phase diagram, and vice versa.  Finally, we restrict 
the analysis to the case $\rho_e=\rho_f$ and plot the phase diagram as a
function of the total filling for each chosen value of $U$.  This computational
algorithm is illustrated schematically in Fig.~\ref{fig: flowdiagram}.

We find that of the initial 23,755 candidate phases, only 111 can be found in
the ground-state phase diagram for the values of $U$ that we considered.  Any 
phase energetically excluded from appearing in the restricted phase diagram
must also be excluded from the complete phase diagram.  What we do not know
is how our computed phase diagram will change as more candidate phases are
introduced (although the majority of these additional phases also won't appear
in the phase diagram).

\section{Results}

The different phases that are stabilized in our restricted phase diagram can 
be grouped into different families that represent different types of geometric
arrangements of the localized electrons.  Unfortunately, there is no way to
rigorously categorize these phases, so the grouping we have chosen arises
in part from our personal taste in determining which phases appear most
similar.  Nevertheless, the groupings we have made are in some sense
``obvious'', and we believe the analysis presented here is a useful way
to categorize and summarize the data.  We will concentrate on describing
different kinds of striped phases that are present in the phase
diagram and we will motivate some of the physical principles behind their
appearance in the phase diagram.

We separate the different stable phases into 10 different groups.  Every
stable phase is labeled by a number and depicted in Fig.~\ref{fig: config}.
The small dots indicate the absence of a localized electron, while the large
dots indicate the positions of the localized electrons.  In the lower
left corner, we shade in the unit cell of the configuration and we show with
the two solid lines the translation vectors of the unit cell that allow the 
square lattice to be tiled by the unit cell.  The different families of
configurations are as follows: (i) \textit{the empty lattice} ($\rho_e\ne 0$
and $\rho_f=0$) denoted E which contains no localized electrons
(configuration 1); 
(ii) \textit{the full lattice} ($\rho_e=0$ and $\rho_f=1$) denoted F which
contains a localized electron at each site (configuration 2); 
(iii) \textit{the checkerboard
phase} ($\rho_e=\rho_f=1/2$) denoted Ch which has the localized electrons
occupying the A sublattice only of the square lattice in a checkerboard
arrangement (configuration 3); 
(iv) \textit{diagonal non-neutral stripe phases} ($\rho_e\ne 1-\rho_f$)
denoted DS
which consist of diagonal checkerboard phases separated by empty diagonal
stripes of slope 1 (configuration 4);
(v) \textit{axial non-neutral checkerboard stripes} ($\rho_e\ne 1-\rho_f$)
denoted AChS which consist of checkerboard regions arranged in stripes oriented 
parallel to the $x$-axis and separated by empty stripes with slope 0
(configurations 5--10);
(vi) \textit{diagonal neutral stripe phases} ($\rho_e=1-\rho_f$) denoted DNS
which consist of localized electrons arranged in the checkerboard phase 
\and separated by fully occupied striped regions of slope 1, or equivalently,
checkerboard phases with diagonal antiphase boundaries (configurations 11--19);
(vii) \textit{axial non-neutral stripe phases} ($\rho_e\ne 1-\rho_f$) 
denoted AS which consist of fully occupied vertical (or horizontal)
stripes separated by empty stripes, which are translationally invariant in the 
vertical (or horizontal) direction (configurations 20--54);
(viii) \textit{neutral phases} ($\rho_e=1-\rho_f$) denoted N which consist
of neutral phases in an arrangement that does not look like any simple
stripe phase (some neutral phases can be described in a stripe picture,
such as configuration 61 which has a slope 1/3 empty lattice stripe, but
we prefer to refer to them as non-stripe phases) (configurations 55--70);
(ix) \textit{four-molecule phases} ($\rho_e\ne 1-\rho_f$) denoted 4M which 
can be described as a ``bound'' four-molecule square of empty sites tiled
inside an occupied lattice framework (configurations 71--74);
(x) \textit{two-dimensional non-neutral phases} ($\rho_e\ne 1-\rho_f$)
denoted 2D which consist of phases with the localized electrons arranged in a 
fashion that is not stripe-like and requires a two-dimensional unit cell
to describe them (once again, some phases like configuration 75 could
be described as a slope 3/2 stripe, but appears to us more like a 2D phase)
(configurations 75--111).

Generically, we find the canonical phase diagram does not contain pure
phases from one of the 111 stable phases, but rather forms mixtures of
two or three periodic phases, or one or two periodic phases and the 
empty lattice (which is often needed to get the conduction-electron filling
correct in the mixture).  When we are doped sufficiently far from half
filling, we are in the segregated phase, which is a mixture of the E
and F phases.

We consider 5 different values of $U$ in our computations: $U=1$, 2, 4, 6, and
8.  The phase diagram is quite complex, with many of the different 111 phases
appearing for different values of $U$.  We summarize which phases appear in
Table~\ref{table: phases}.  

We begin our discussion with the weak-coupling value $U=1$ where 50
phases appear.  The phase diagram is summarized in Fig.~\ref{fig: u=1}.
We use a solid line to indicate the region of the particle density
where a particular phase appears in the ground state (either as a pure
phase or as a mixture).  The phases that appear in a mixture at a given density
are found by determining the solid vertical lines that intersect a horizontal 
line drawn to pass through the given particle density.  The phase diagram has
shading included to separate the regions of the different categories of phases.
The numeric labels are shown to make it easier to determine the actual phases
present in the diagram.
We plot similar phase diagrams for $U=2$ (38 phases), 4 (42 phases), 6
(30 phases), and 8 (25 phases) in 
Figs.~\ref{fig: u=2}--\ref{fig: u=8}, respectively.  A schematic phase diagram
that illustrates the generic features of the phase diagram in the electron
density, interaction-strength plane appears in Fig.~\ref{fig: schematic}.

As can be seen from these figures, the generic phase diagram is quite
complex, and by looking at the different phases in Fig.~\ref{fig: config},
many of the phases have stripe-like structures to them.  To begin our
discussion of these results, we must first recall the rigorous results known
for this model.  When $\rho_e=\rho_f=1/2$, the ground state is the checkerboard
phase (configuration 3) for all $U$.  This can be seen in all of the
phase diagrams plotted.  When $\rho_e=\rho_f\ne 1/2$, the ground state 
becomes the phase separated segregated phase when $U$ is large enough.
So there is a simplification in the phase diagram as we increase $U$,
and the most complex phase diagram appears in the $U\rightarrow 0$ limit.
That limit is also the most difficult computationally, because the differences
in the energies between different configurations also becomes small for
small $U$, and the numerical accuracy must be huge in order to achieve
trustworthy results.  This is why we do not report any phase diagrams with
$U<1$ here.

Looking at the $U=8$ case shown in
Fig.~\ref{fig: u=8}, we see that as we move away from half filling, 
we initially find mixtures between the checkerboard phase, other diagonal
stripe phases, and the empty lattice.  When we examine the structure factors
associated with the diagonal stripe phases, we find that they tend to have more
weight along the Brillouin zone diagonal than elsewhere.  Hence, these diagonal
stripe phases are being stabilized by a ``near-nesting'' instability of the
noninteracting Fermi surface, and the overall mixtures are required to 
maintain the average fillings of the conduction and localized electrons.
As we move farther from half filling, the checkerboard phase disappears from the
mixtures, and then a series of neutral phases enter the mix which retain
some appearance of diagonal stripes, but with more and more ``defects''
to the stripes that make them look more two-dimensional.  We find the localized
electron density of these phases increases as we reduce the total filling,
which is what we expect as we move toward the segregated phase which involves
a mixture of the E and F configurations.  Note that the formation of many
different stripe phases, occurs without needing the long-range Coulomb
interaction to oppose the tendency towards phase separation, when we are
close to half filling.  Indeed, the ground state is often a phase separated
mixture, but it is a mixture of stripe-like phases, which occur automatically,
without the need to add any other physics to the system.  This regime, is the
closest to the Kivelson-Emery picture, but we see it has more complex behavior
than what they envisioned when they examined the Hubbard model.

Moving on to the $U=6$ case in Fig.~\ref{fig: u=6}, we find a significant
change in the phase diagram.  The grouping of diagonal stripes near half
filling disappears and we instead find the ground state to initially be a 
mixture between the checkerboard phase, a truly two-dimensional screen-like
phase (configuration 108) and the empty lattice.  Here, if we include a 
long-range Coulomb interaction, we would likely form diagonal stripes, but
the mixture would be more complicated because it would include this 
screen-like structure as well.  As we dope further away, we see a smaller 
number of the neutral phases, which look somewhat like diagonal stripes with
a large number of defects in them, and then we go to a very different class
of mixtures, dominated by the presence of the axial stripe phase in 
configuration 33.  As that phase becomes destabilized, we find a cascade of many
other axial stripes entering, before the segregated phase takes over.  This
transition from diagonal stripes to axial stripes as a function of the electron
filling, also occurs because of a ``near-nesting'' effect.  The structure
factors of the axial stripe phases are peaked predominantly along the zone edge,
and as we dope further from half filling, this is where nesting is more
likely to occur. The cascade of stable phases that enter after configuration
33 is destabilized, have a progression of the peaks in their structure factor
moving towards the zone center, which is also expected, since they are
progressively heading towards the segregated phase.
A similar kind of transition from diagonal stripes
to axial stripes is seen in the Hubbard model studies, with the critical
density lying near 0.375, as we see here too.

By the time we decrease to $U=4$ shown in Fig.~\ref{fig: u=4}, we find
even more interesting behavior.  Now, when we are near half filling, we find
two more configurations, a nonneutral phase (configuration 59) and a 
two-dimensional phase (configuration 109) joining with the checkerboard
phase and configuration 108 in the initial mixtures. Each of these phases
looks like a ``square-lattice screen'' with differing size ``holes'' in the
screen.  These two-dimensional structures are not stripe-like and it would
be interesting to see if they could appear in the Hubbard model.  As we dope
further away, we enter the axial stripe region, now dominated by configuration
20 first, then there is a cascade to configuration 33, then a cascade to the 
segregated phase.  This value of $U$ is a truly intermediate value, where 
many different mechanisms for ordering are present and the system can change
very rapidly in response to a modification in the density.

As $U=2$ (Fig.~\ref{fig: u=2}), we see more modifications in the phase diagram.
Now we see other diagonal stripe phases mixing with the checkerboard phase
near half filling.  This region would correspond to the Scalapino-White
regime, where the stripe formation is driven more by interplays between
the kinetic and potential energies and nesting effects (driven by charge
fluctuations in the FK model and spin fluctuations in the Hubbard model).
In addition, a much larger number of the 2D phases enter also close to
half filling, illustrating the prevalence of these ``screen-like''
phases as well.  The axial stripes also enter as we dope further away
from half filling, but the configurations 20 and 33 are not nearly as stable
as they are for slightly larger $U$.  Here, we see the four-molecule phases
being stabilized
just before the system phase separates into the segregated phase.

Finally, for $U=1$, shown in Fig.~\ref{fig: u=1}, the predominance of the
diagonal stripes, near half filling increases now supplemented by the
axial checkerboard stripes, but then there is a plethora
of different 2D phases that also enter as the system is doped somewhat farther
from half filling, then we see a similar evolution, first to AS and then
to 4M phases before the segregated phase.  Here there is a tremendous 
complexity to the phase diagram, with many different mixtures being present due
to the competition between kinetic energy and potential energy minimization
brought about by the many-body aspects of the problem.

The general picture, illustrated schematically in Fig.~\ref{fig: schematic},
now emerges: near half filling, we often find diagonal stripes and screen-like
two-dimensional phases, then a rapid transition to the segregated phase
for large $U$.  As $U$ is reduced, we can dope farther away from half filling
before segregating, which allows many other phases to enter.  In particular,
there is a large region of stability for axial stripes, and as $U$ is
reduced further, we see the emergence of axial checkerboard stripes close to
half filling, near the diagonal stripes, and four-molecule phases appearing
near the segregation boundary.

\section{Conclusions}

In this manuscript we have numerically studied how the FK model makes the
transition from the checkerboard phase at half filling to the segregated
phase as the density is lowered.  Since these two phases are very different
from one another, there are many different pathways that one might imagine
the system to take in making this crossover.  Indeed we find that the
pathway varies dramatically as a function of $U$.  For large $U$, we have a 
relatively simple transition between diagonal stripe-like phases which
become more two-dimensional as the localized electron density increases, until
the system gives way to the segregated phase.  As $U$ is lowered, we first
see two-dimensional-``screen''-like phases enter, then we see axial stripes
emerge, followed by four-molecule phases and axial checkerboard stripes.
The complexity of the phase diagram greatly increases as the interaction 
strength decreases.  

It is interesting to ask how we might expect these results to change if we
allowed more configurations into our restricted phase diagram.  We don't know
this answer in particular, but we do know, that of the 23,755 candidate
phases only a small fraction (111 or 0.5\%) 
appear in the phase diagram, so we don't
expect too many additional phases to appear.

Another interesting question to ask is how do these results for the FK model
shed light on the stripe-formation problem in the Hubbard model. By continuity,
we expect these results not to change too dramatically as we turn on a small
hopping for the localized electrons (although now we must summarize our results
in terms of correlation functions for the two kinds of electrons, since both
are now mobile). But we also know for many fillings, there will be a
``phase transition'' as a function of the hopping, since the ground state
of the Hubbard model is not ferromagnetic for all fillings and large $U$
(which is what the segregated phase maps to in the Hubbard model).  The
results are likely to be closer to what happens in the Hubbard model close to 
half filling, because the analogue of the antiferromagnetic phase is the 
checkerboard phase, and that is present for all $U$ in the Hubbard model
at $T=0$.  In general, we also feel that the FK model phase diagram must
be more complicated than the Hubbard model phase diagram because of the
mobility of both electrons in the latter.
We feel one of the most important results of this work is
that there may be two-dimensional phases that are not stripe like that
form ground-state configurations for some values of the filling in the
Hubbard model, and such configurations will be worthwhile to investigate
with the numerical techniques that currently exist.

In conclusion, we are delighted to be able to shed some light on the
interesting question for the FK model of how one makes a transition from
the checkerboard phase at half filling to the segregated phase away
from half filling.  Since Elliott Lieb has had an important impact in 
proving the stabilization of these two phases, we find it fitting to ask the
questions about how the two phases inter-relate.  Perhaps these numerical
calculations can further inspire new rigorous work that helps to identify
the pathway between these two phases.

\section{Acknowledgments}

R.L. and G.B. acknowledge support from the Polish State Committee for Scientific
Research (KBN) under Grant No. 2P03B 131 19 and J.K.F. acknowledges support from
the NSF under grant No. DMR-0210717. We also acknowledge support from
Georgetown University for a travel grant in the fall of 2001.

\bibliography{fk_stripes.bib}

\newpage

\center{\large \textbf{Tables}}

\begin{table}[h]
\caption{\label{table: phases} Summary of the stability of different phases for
the five different values of $U$ where we performed calculations (1, 2, 4, 6, 
and 8). Each column shows the phases that appear in the phase diagram
for a given value of $U$. The numbers correspond to the labels in 
Fig.~\ref{fig: config}.}
\begin{ruledtabular}
\begin{tabular}{llllll}
Phase category&$U=1$&$U=2$&$U=4$&$U=6$&$U=8$\\
\colrule
E&1&1&1&1&1\\
F&2&2&2&2&2\\
Ch&3&3&3&3&3\\
DS&4\\
AChS&5-10\\
DNS& &11-14& & &14-19\\
AS&20, 26-33&20, 28-40&20-54&33-52&\\
N& & & 59& 59-64& 55-70\\
4M&71-73&73-74&\\
2D&75-78, 80-82&79, 82-83,&83,&108\\
  &84-92, 94-100,&87-88, 93,&108-109\\
  &103-107&98-102, 105-106&\\
  & &110-111\\
\end{tabular}
\end{ruledtabular}
\end{table}

\newpage

\center{\large \textbf{Figure Captions}}
{
\begin{figure}[f]
\epsfysize=0.05in
\caption{\label{fig: bands} Bandstructure along the irreducible wedge
of the square lattice Brillouin zone for the truly two-dimensional
configuration numbered 108 and depicted in Fig.~\ref{fig: config}.
In panel (a) we plot the band structure 
and the density of states for $U=2$.  In panel (b) we show
the same for $U=4$.  Note how there is less band overlap as $U$ increases.}
\end{figure}
\begin{figure}[f]
\epsfysize=0.05in
\caption{\label{fig: flowdiagram} Flow chart that illustrates the algorithm
employed to calculate the phase diagram of the Falicov-Kimball model. Note
that of the 23,755 candidate phases, only 111 appear in the restricted
phase diagram.}
\end{figure}
\begin{figure}[f]
\epsfysize=0.05in
\caption{\label{fig: config} Picture of the configurations of the localized
electrons that appear in the restricted phase diagram.  The large dots refer
to sites occupied by localized electrons, and the small circles denote
empty sites.  The shaded region in the lower left corner shows the unit cell,
and the line segment shows the translation vector that is used to tile
the two dimensional plane.  Each of the 111 configurations is assigned
a number, and we also note the size and shape of the unit cell and the
localized electron filling in parenthesis above each panel.}
\end{figure}
\begin{figure}[f]
\epsfysize=0.05in
\caption{\label{fig: u=1} Phase diagram for $U=1$. The solid lines show the
regions of electron density where a particular phase appears (either as
a single phase or as a mixture).  The horizontal axis labels the different
configurations that are present, and the shading helps to distinguish the
different categories of the phases.  The numbers are included as a guide to
make it easier to identify the different stable phases in the diagram.}
\end{figure}
\begin{figure}[f]
\epsfysize=0.05in
\caption{\label{fig: u=2} Phase diagram for $U=2$. The notation is the
same as in Fig.~\ref{fig: u=1}.}
\end{figure}
\begin{figure}[f]
\epsfysize=0.05in
\caption{\label{fig: u=4} Phase diagram for $U=4$. The notation is the
same as in Fig.~\ref{fig: u=1}.}
\end{figure}
\begin{figure}[f]
\epsfysize=0.05in
\caption{\label{fig: u=6} Phase diagram for $U=6$. The notation is the
same as in Fig.~\ref{fig: u=1}.}
\end{figure}
\begin{figure}[f]
\epsfysize=0.05in
\caption{\label{fig: u=8} Phase diagram for $U=8$. The notation is the
same as in Fig.~\ref{fig: u=1}.}
\end{figure}
\begin{figure}[f]
\epsfysize=0.05in
\caption{\label{fig: schematic} Schematic phase diagram which indicates
the different categories of phases that appear in the restricted
phase diagram.}
\end{figure}
}
\newpage

\epsfxsize=4.4in
\epsffile{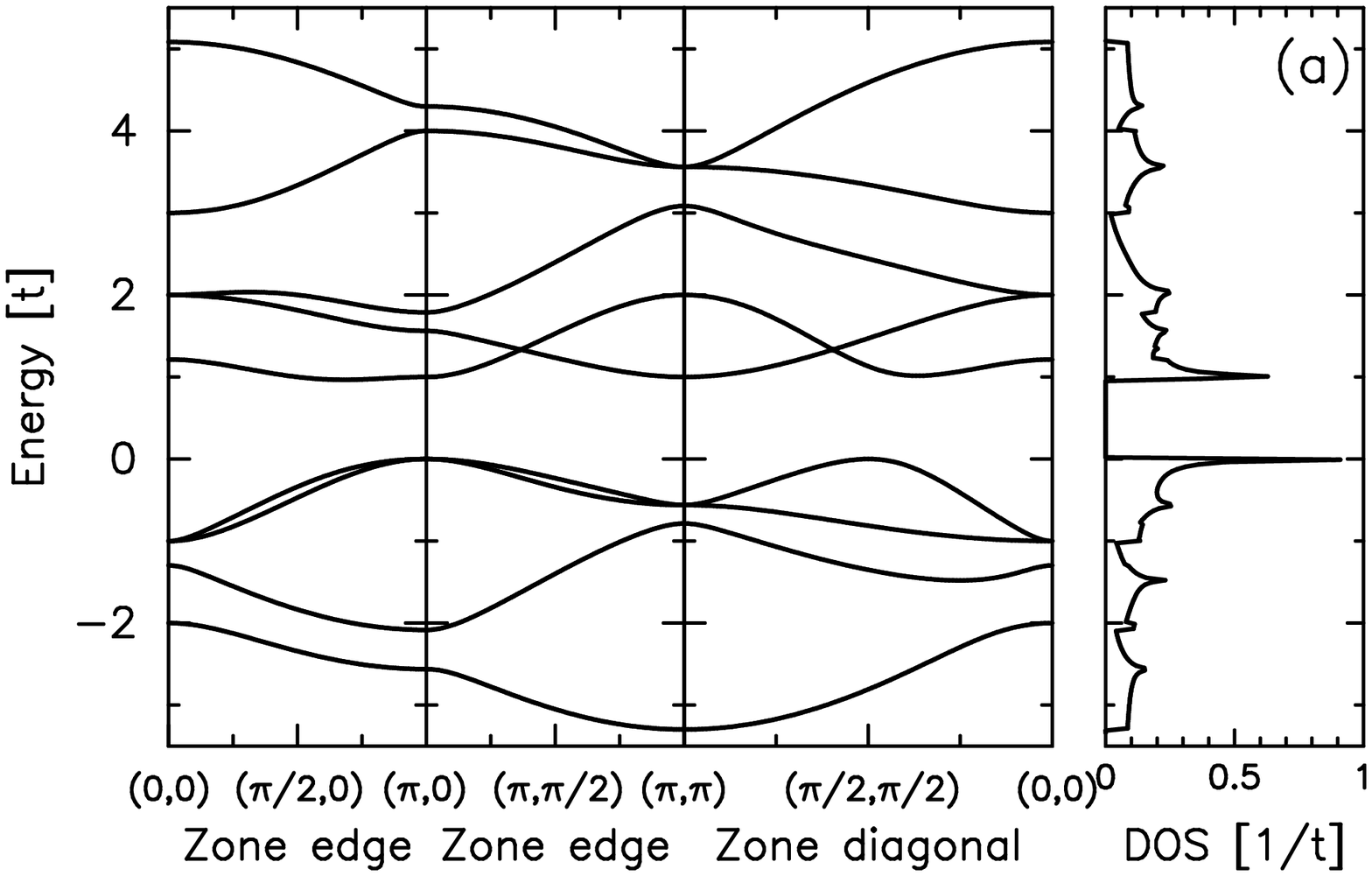}
\epsfxsize=4.4in
\epsffile{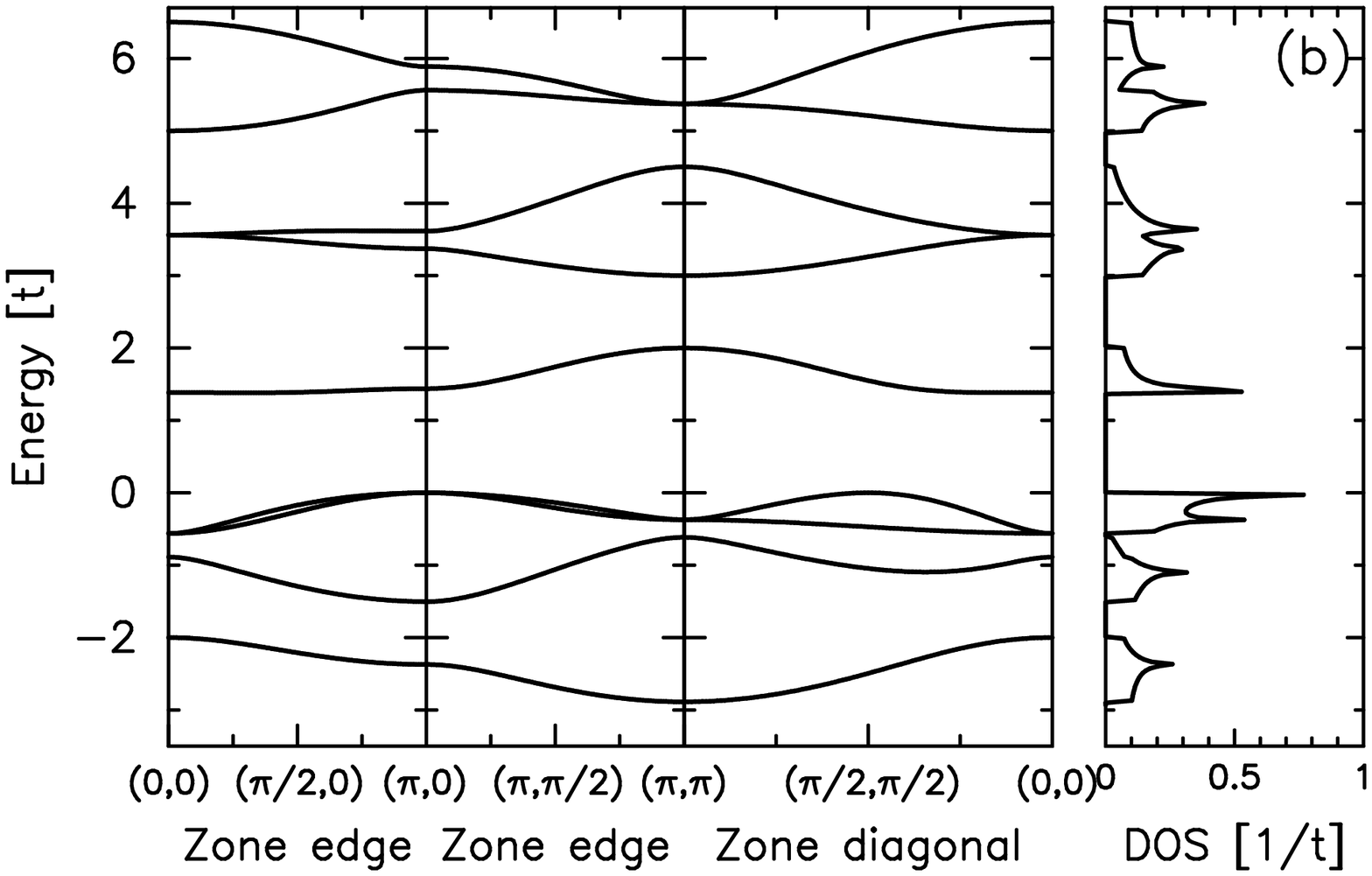}
\vskip 0.5in
Figure 1.

\newpage

\epsfxsize=4.4in
\epsffile{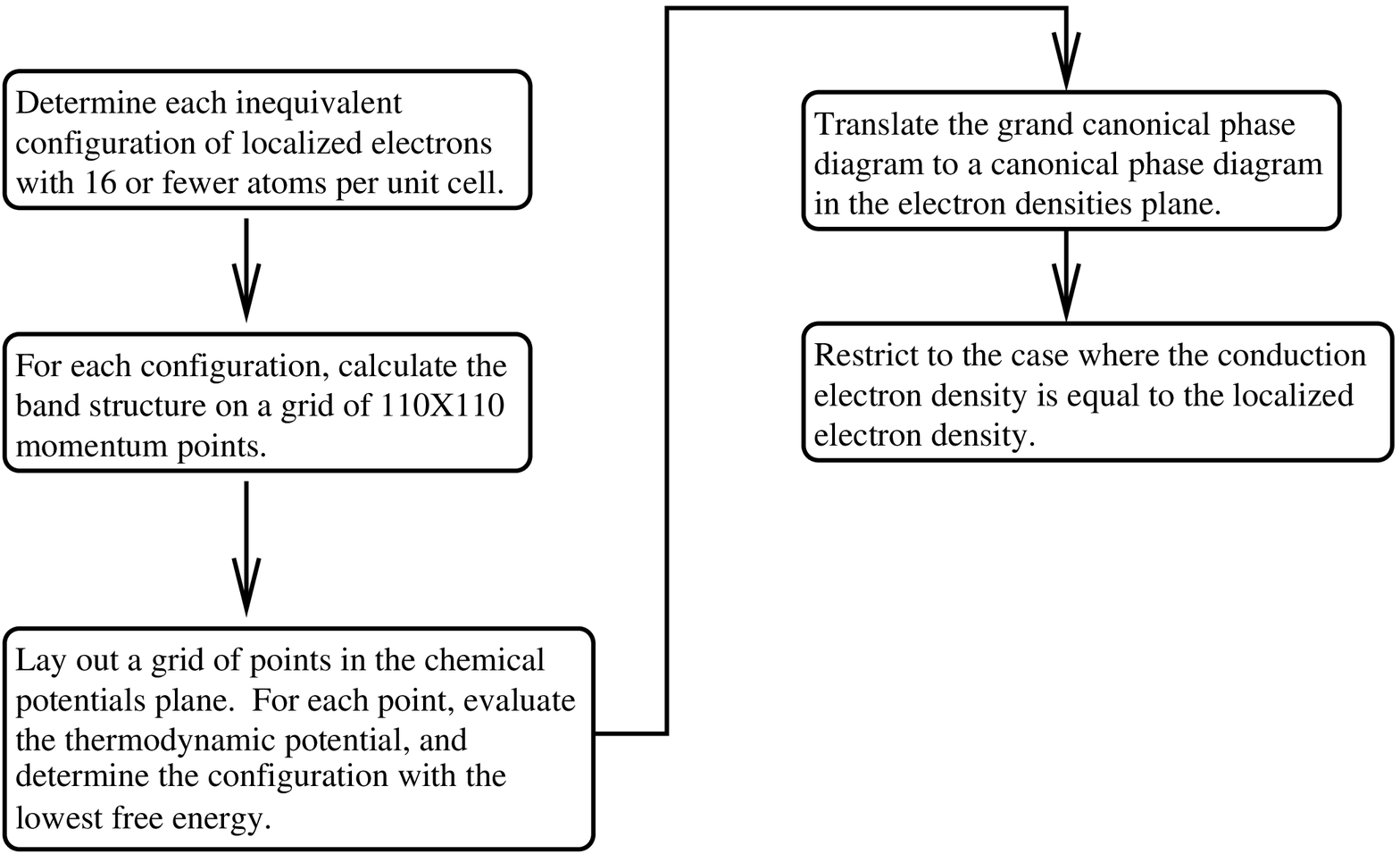}
\vskip 0.5in
Figure 2.

\newpage

\epsfxsize=4.9in
\epsffile{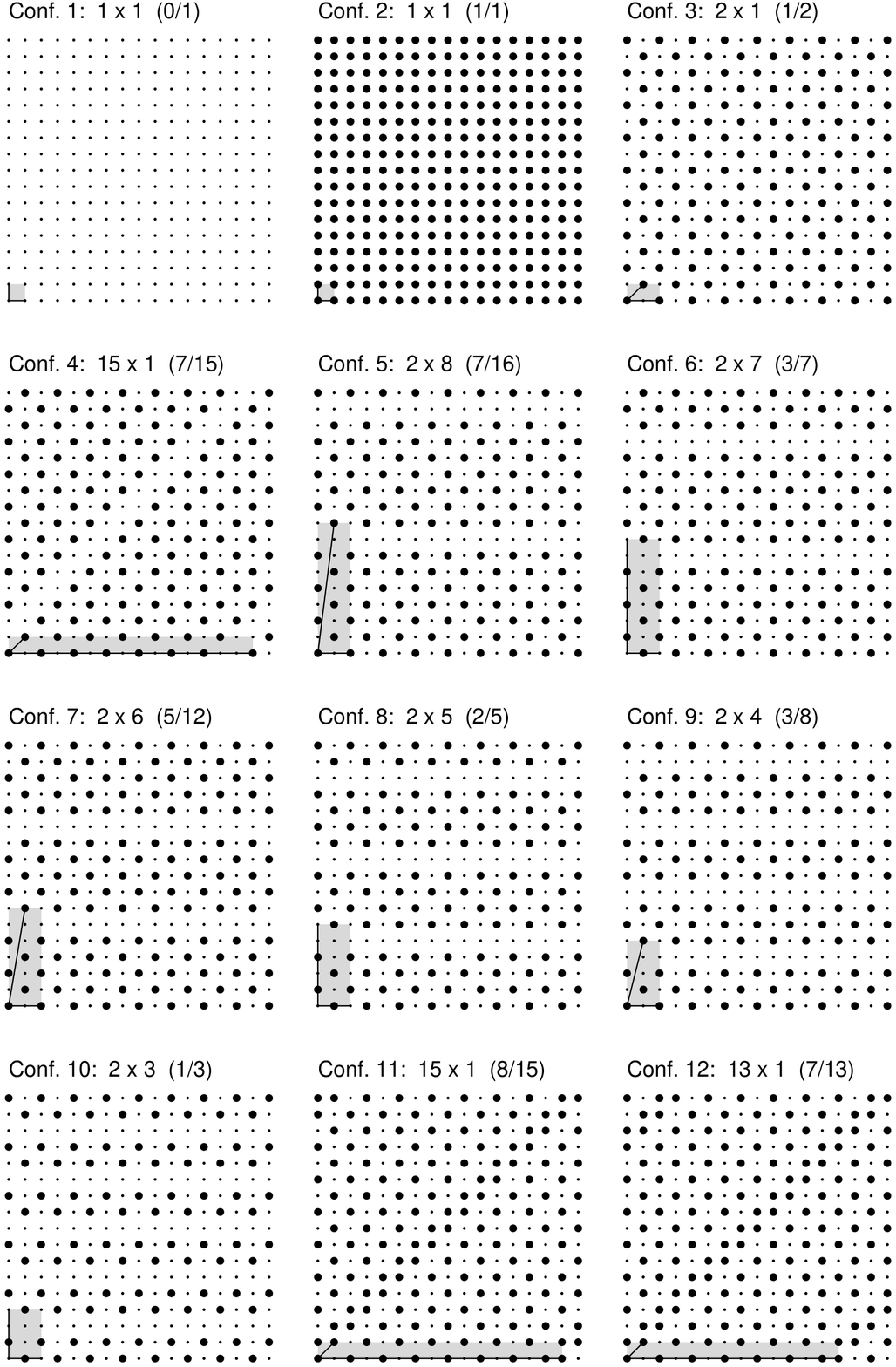}
\epsfxsize=4.9in
\epsffile{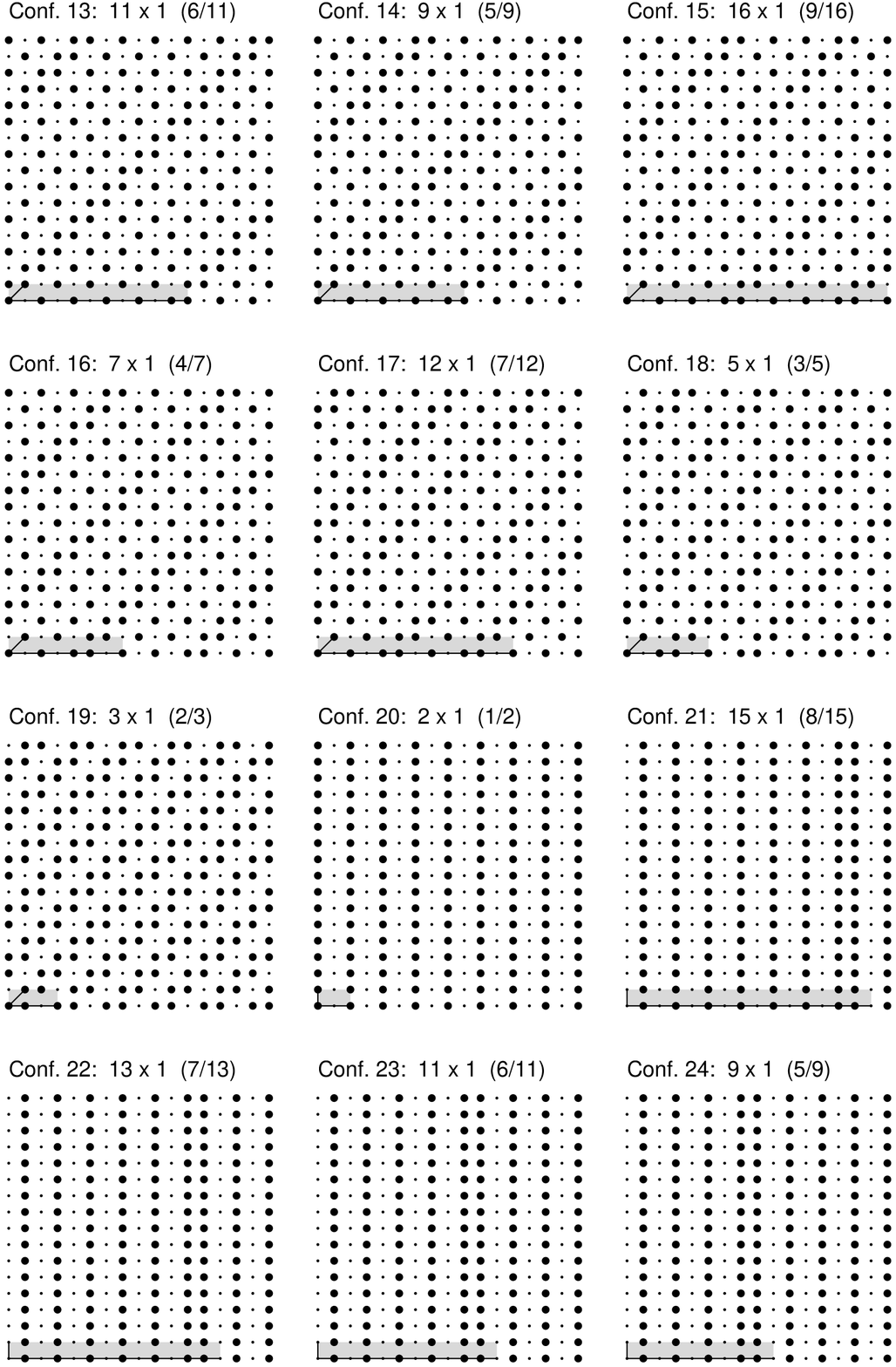}
\epsfxsize=4.9in
\epsffile{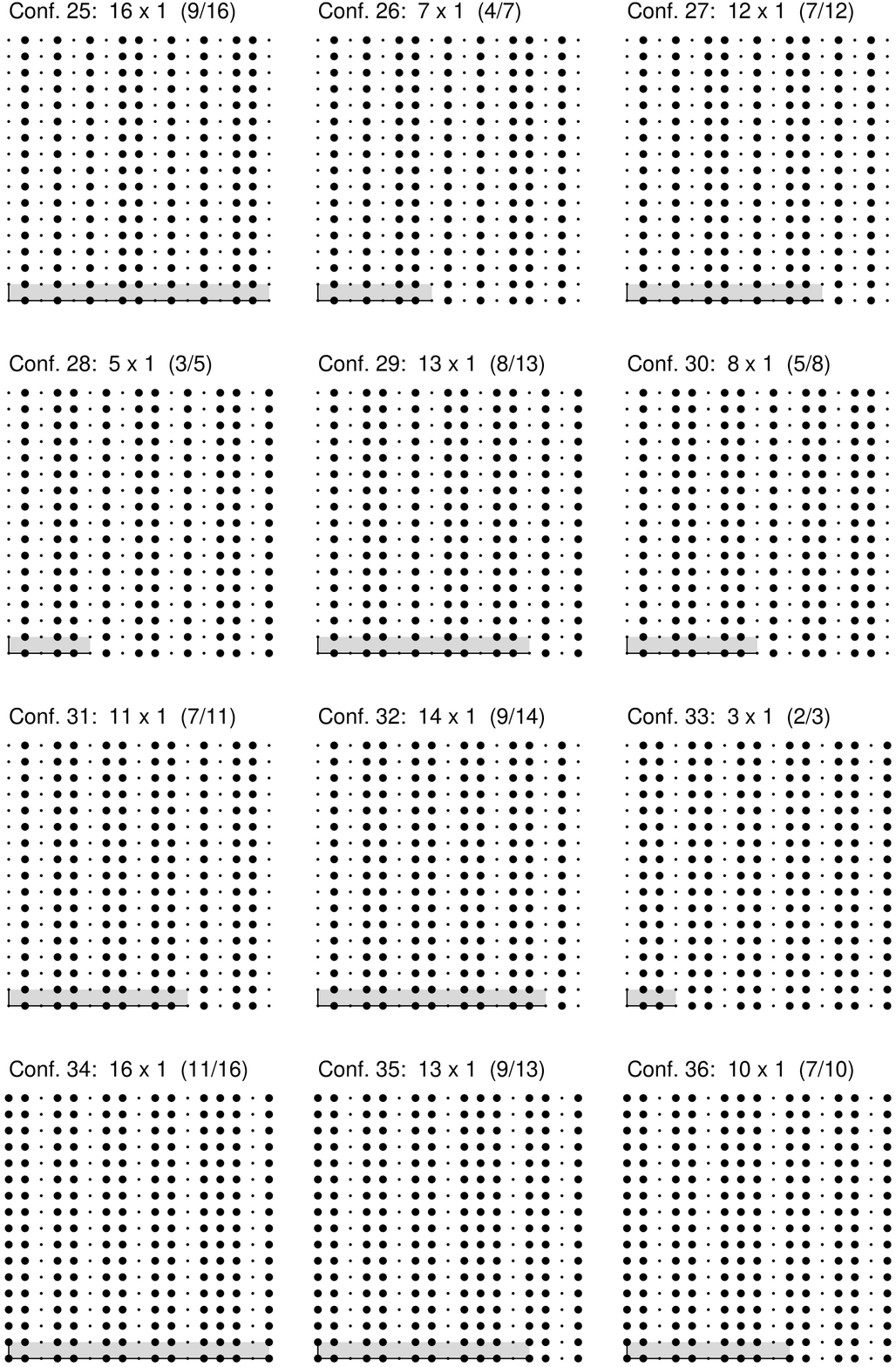}
\epsfxsize=4.9in
\epsffile{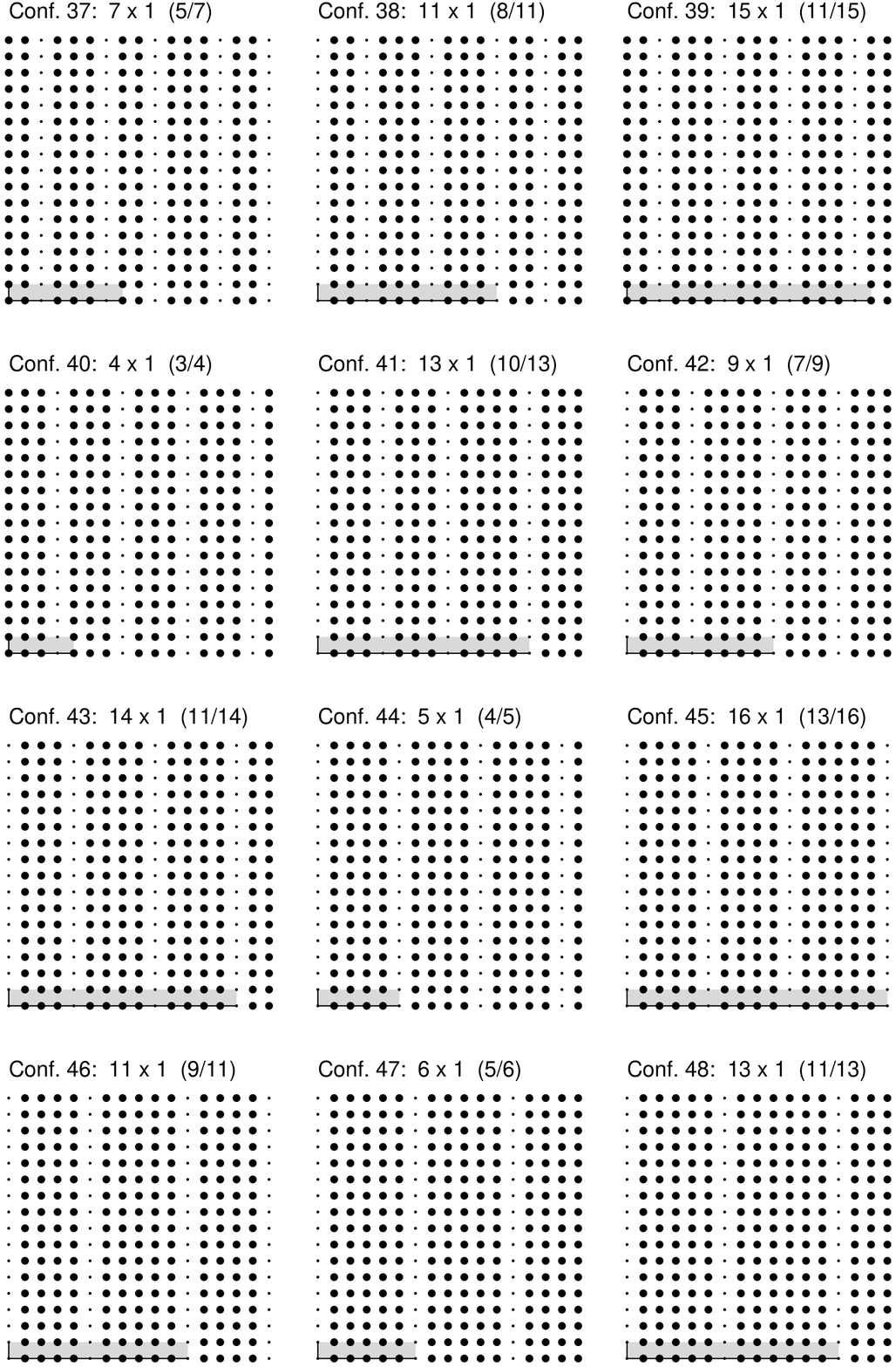}
\epsfxsize=4.9in
\epsffile{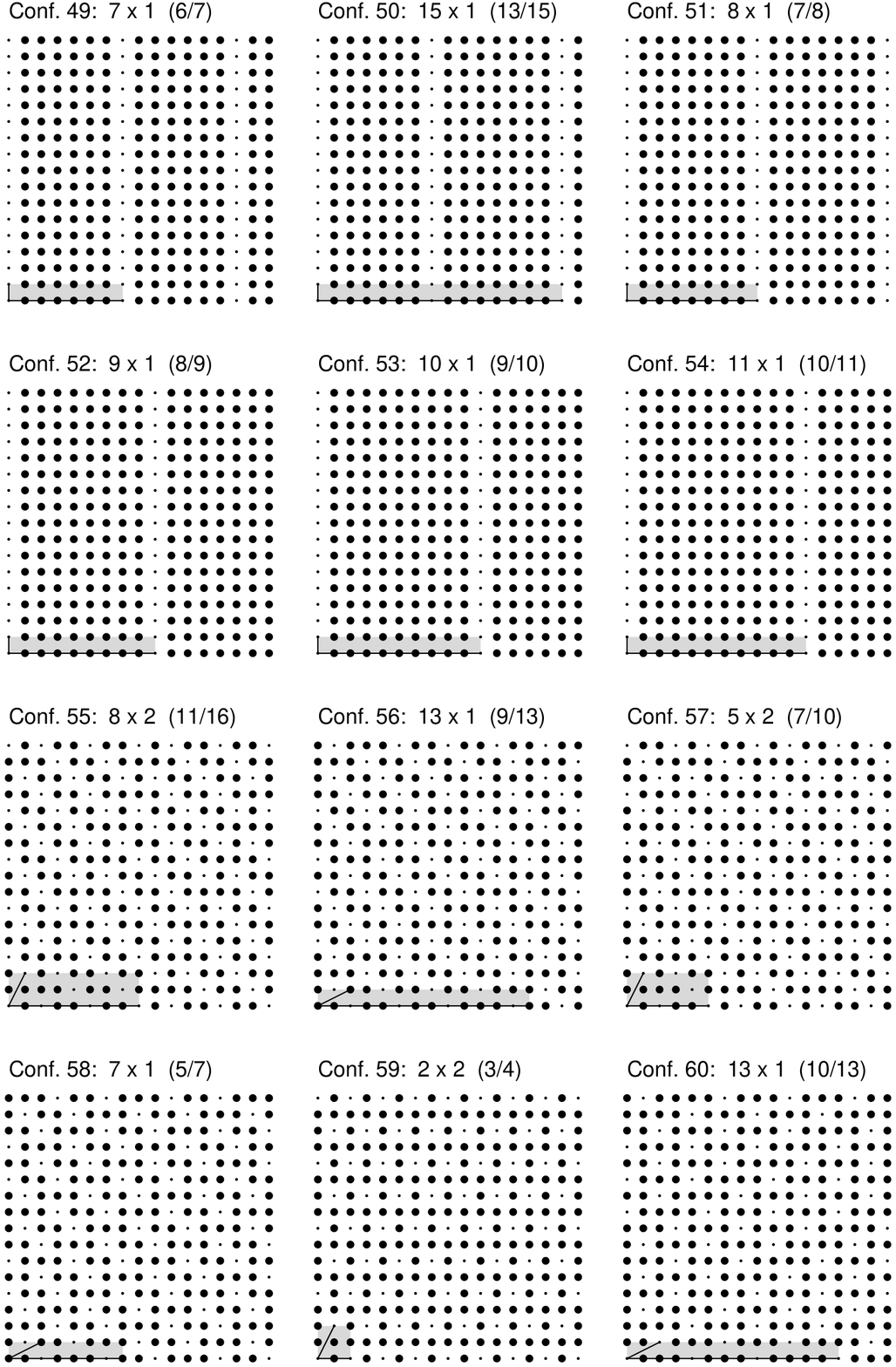}
\epsfxsize=4.9in
\epsffile{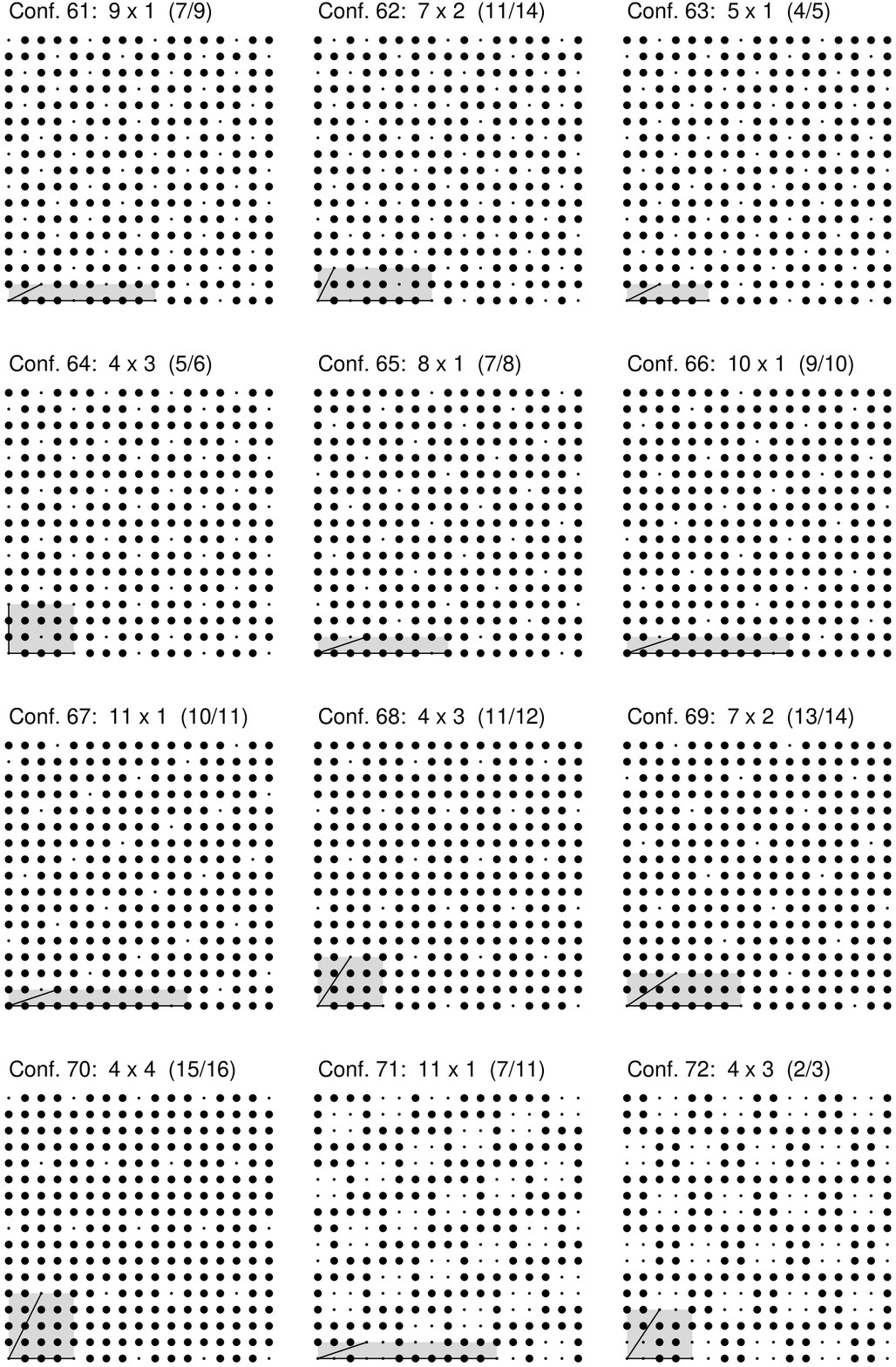}
\epsfxsize=4.9in
\epsffile{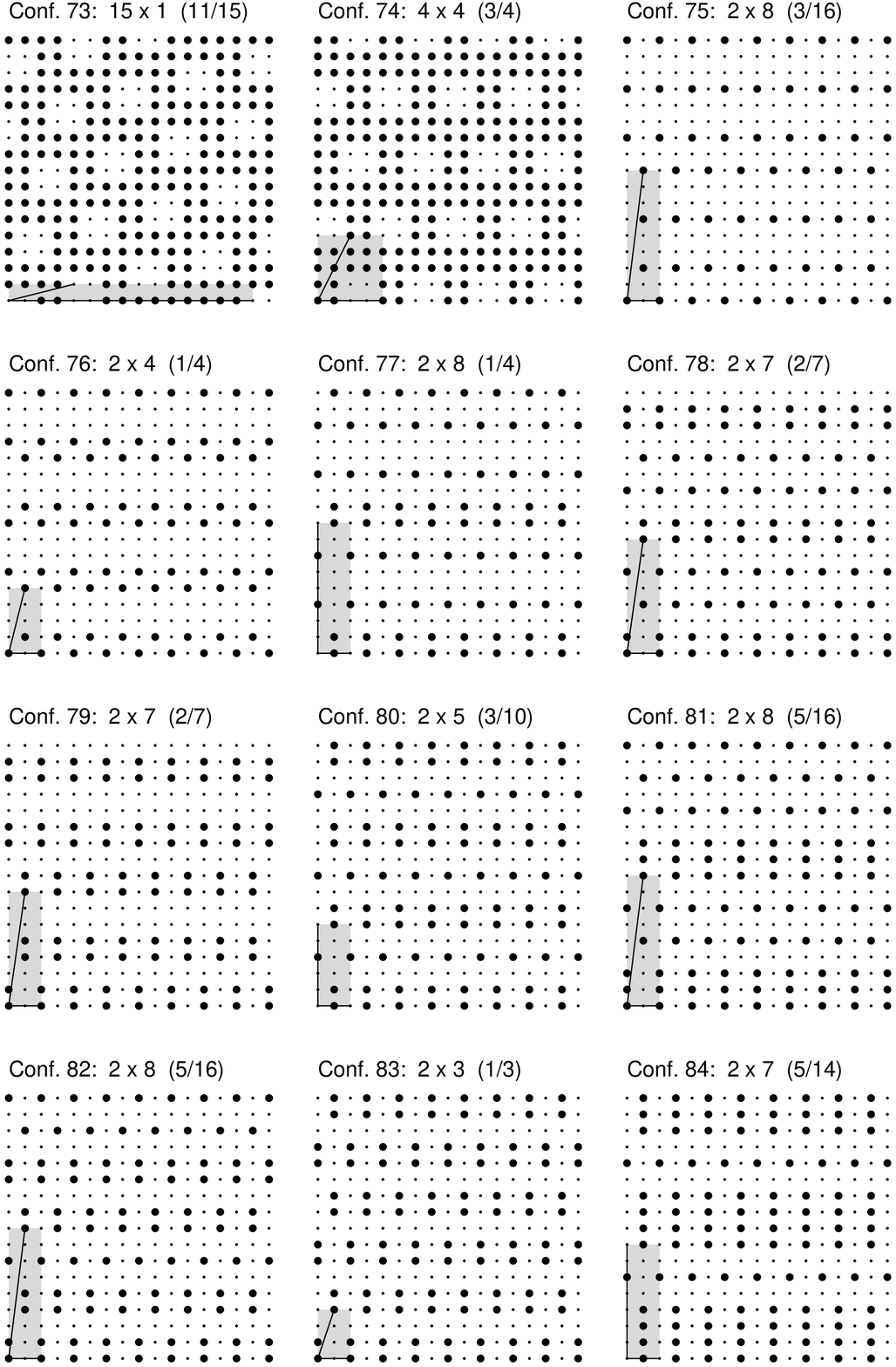}
\epsfxsize=4.9in
\epsffile{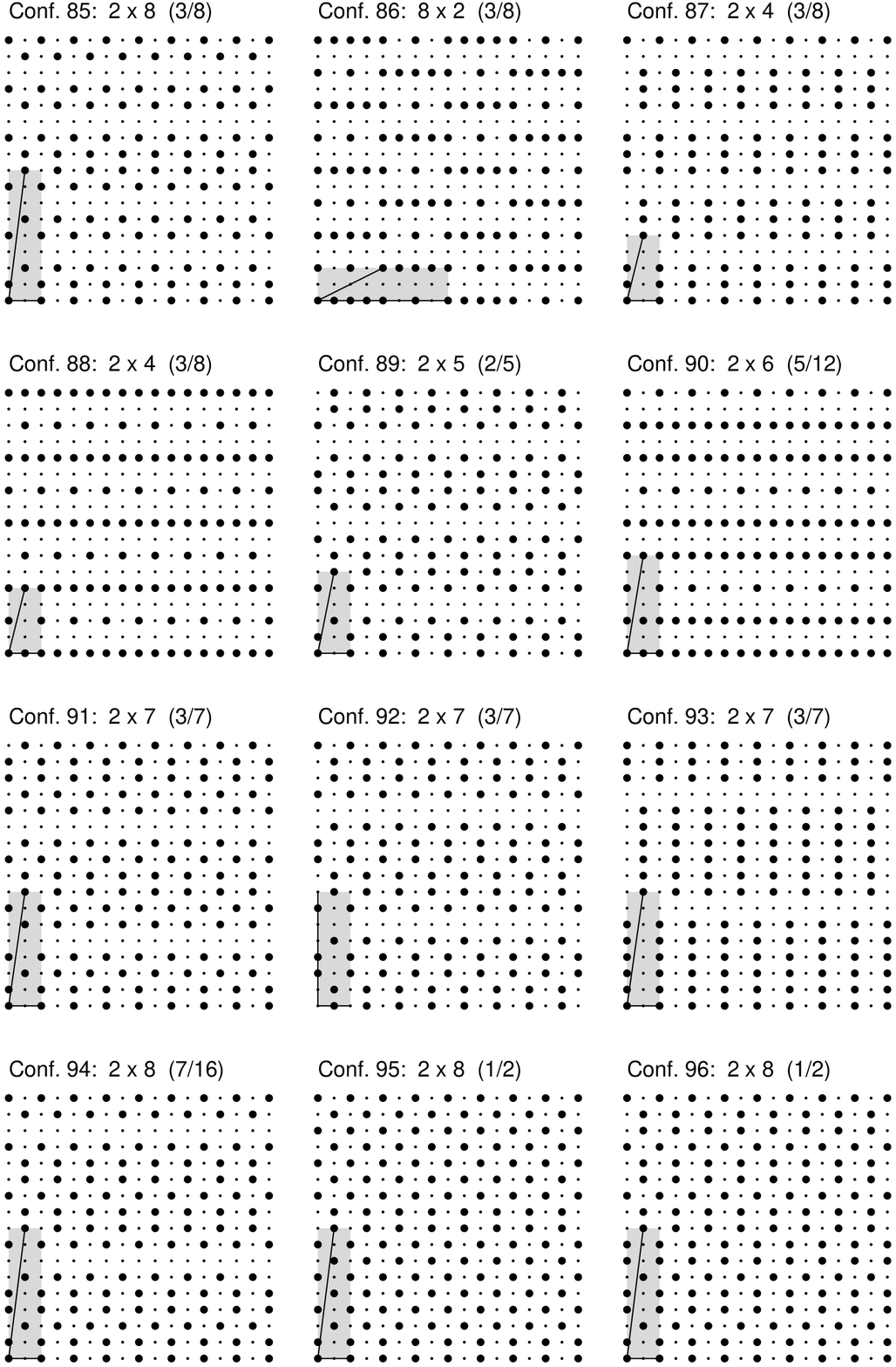}
\epsfxsize=4.9in
\epsffile{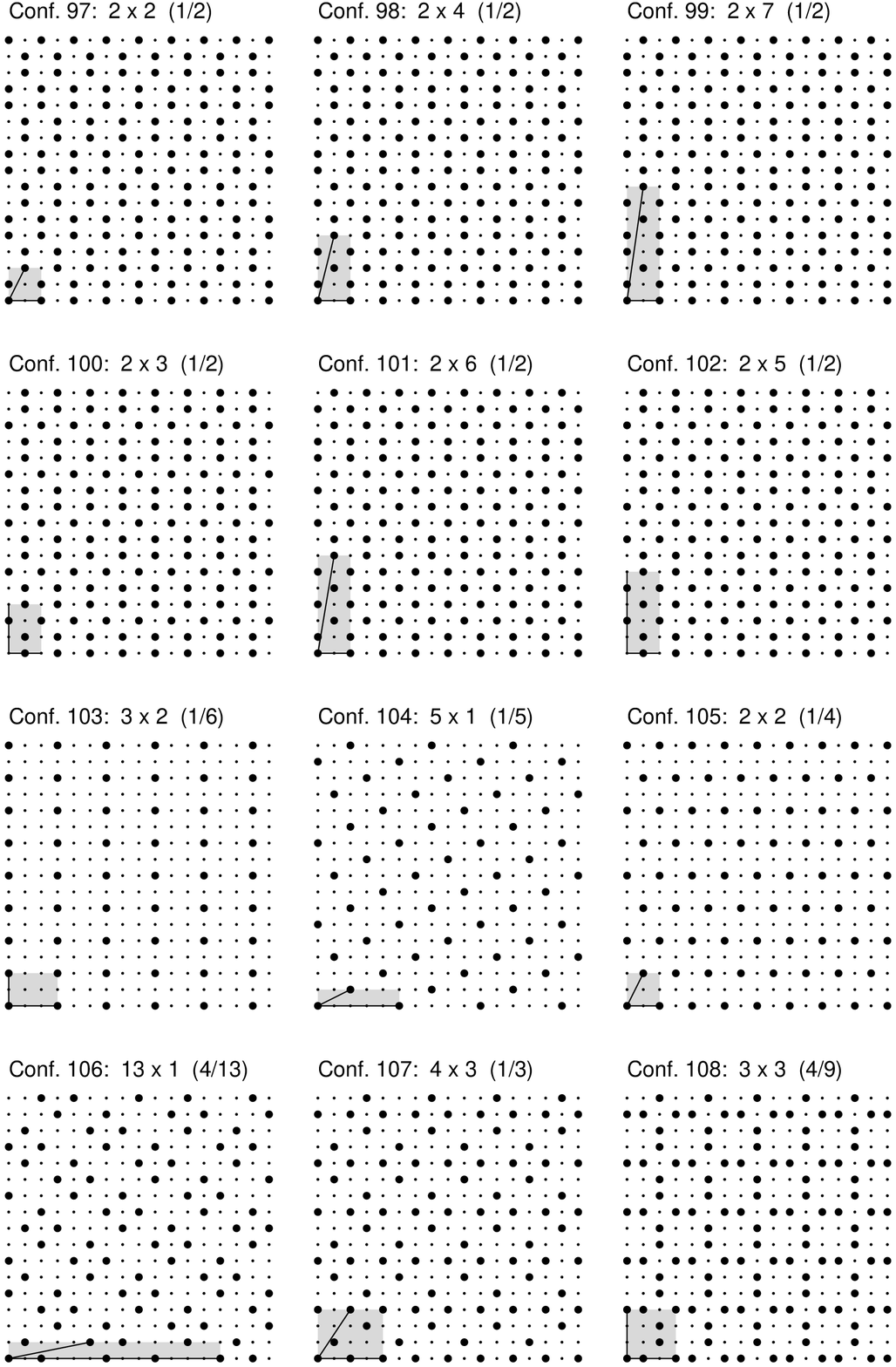}
\epsfxsize=4.9in
\epsffile{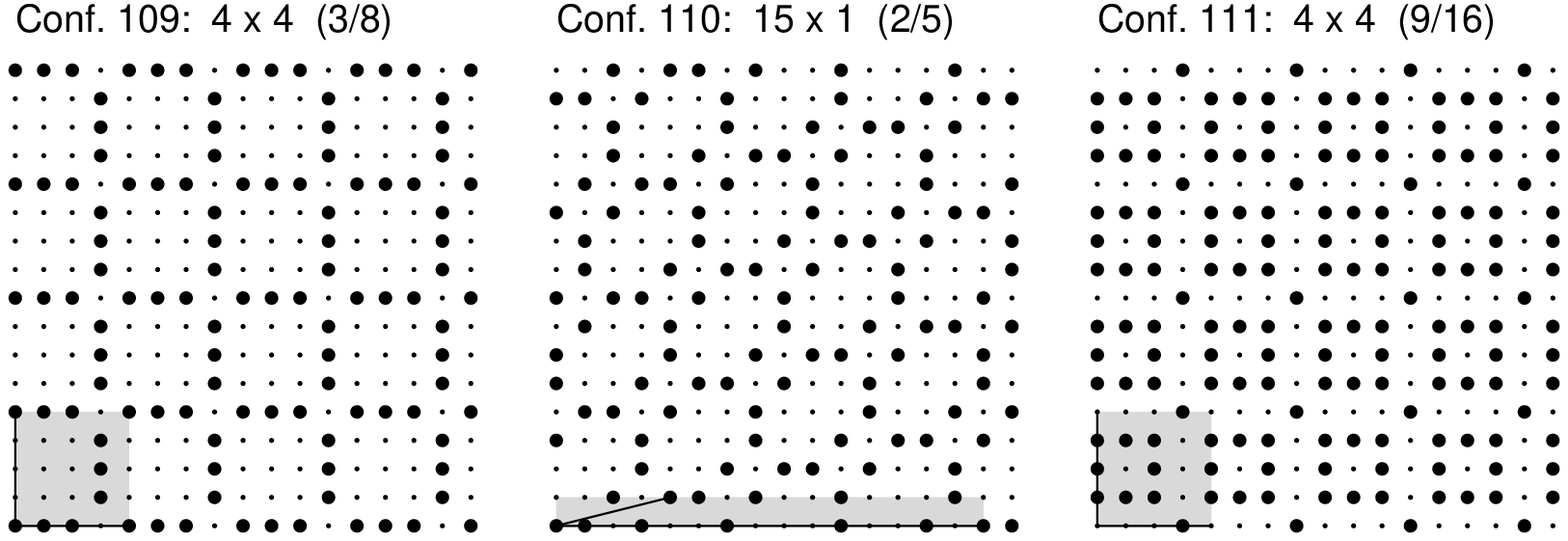}
\vskip 0.5in
Figure 3.

\newpage

\epsfxsize=4.4in
\epsffile{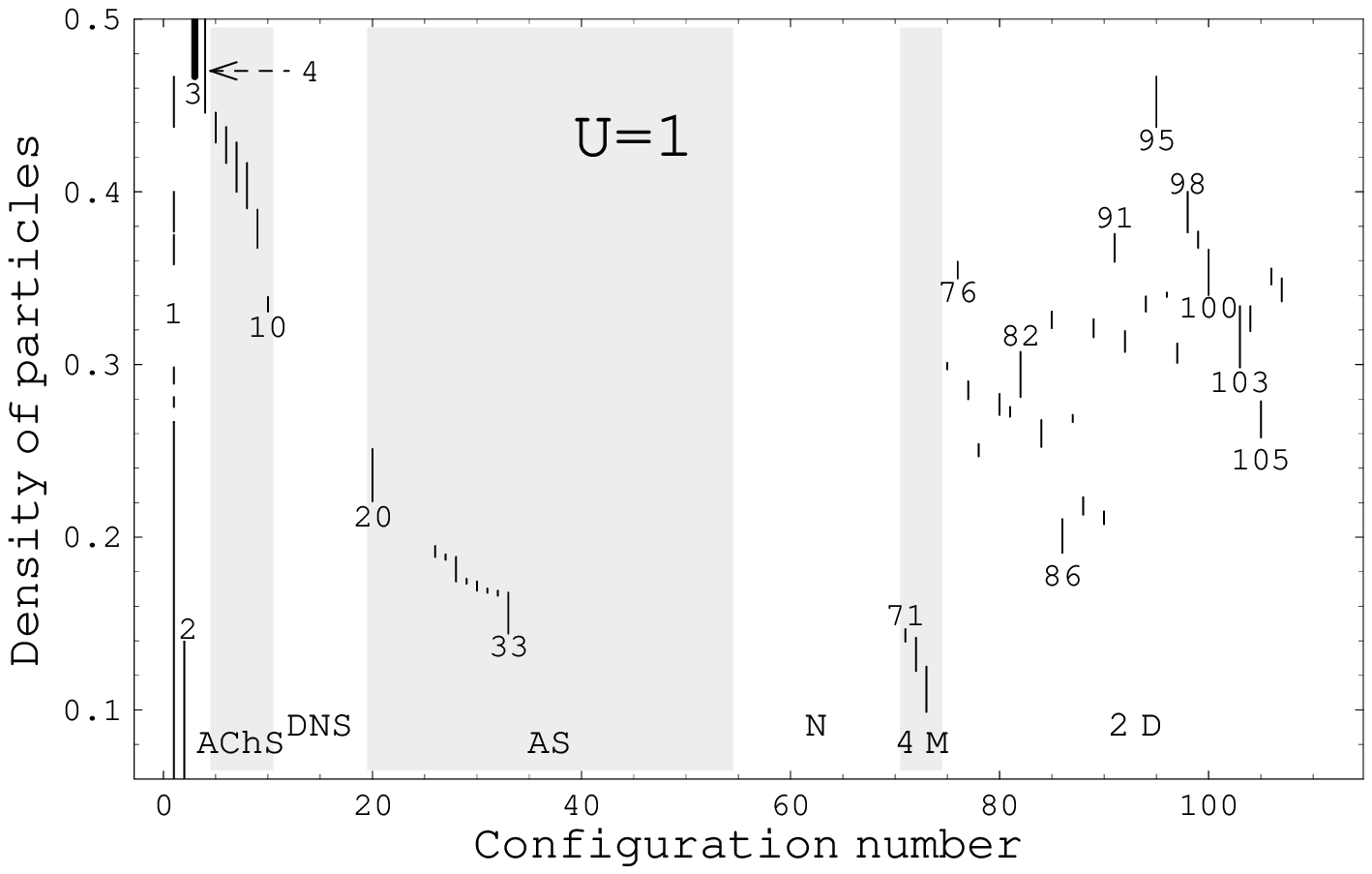}
\vskip 0.5in
Figure 4.

\newpage

\epsfxsize=4.4in
\epsffile{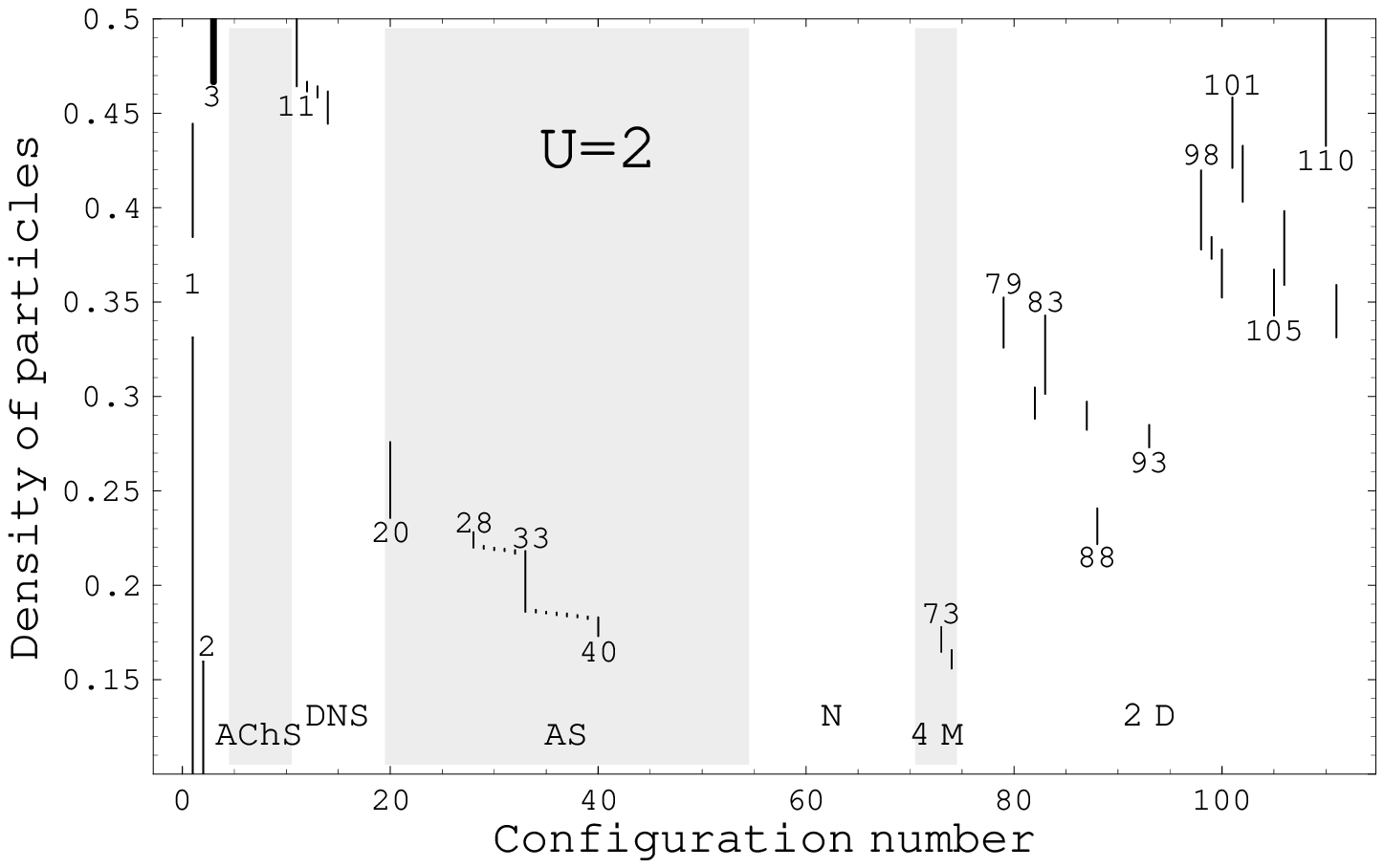}
\vskip 0.5in
Figure 5.

\newpage

\epsfxsize=4.4in
\epsffile{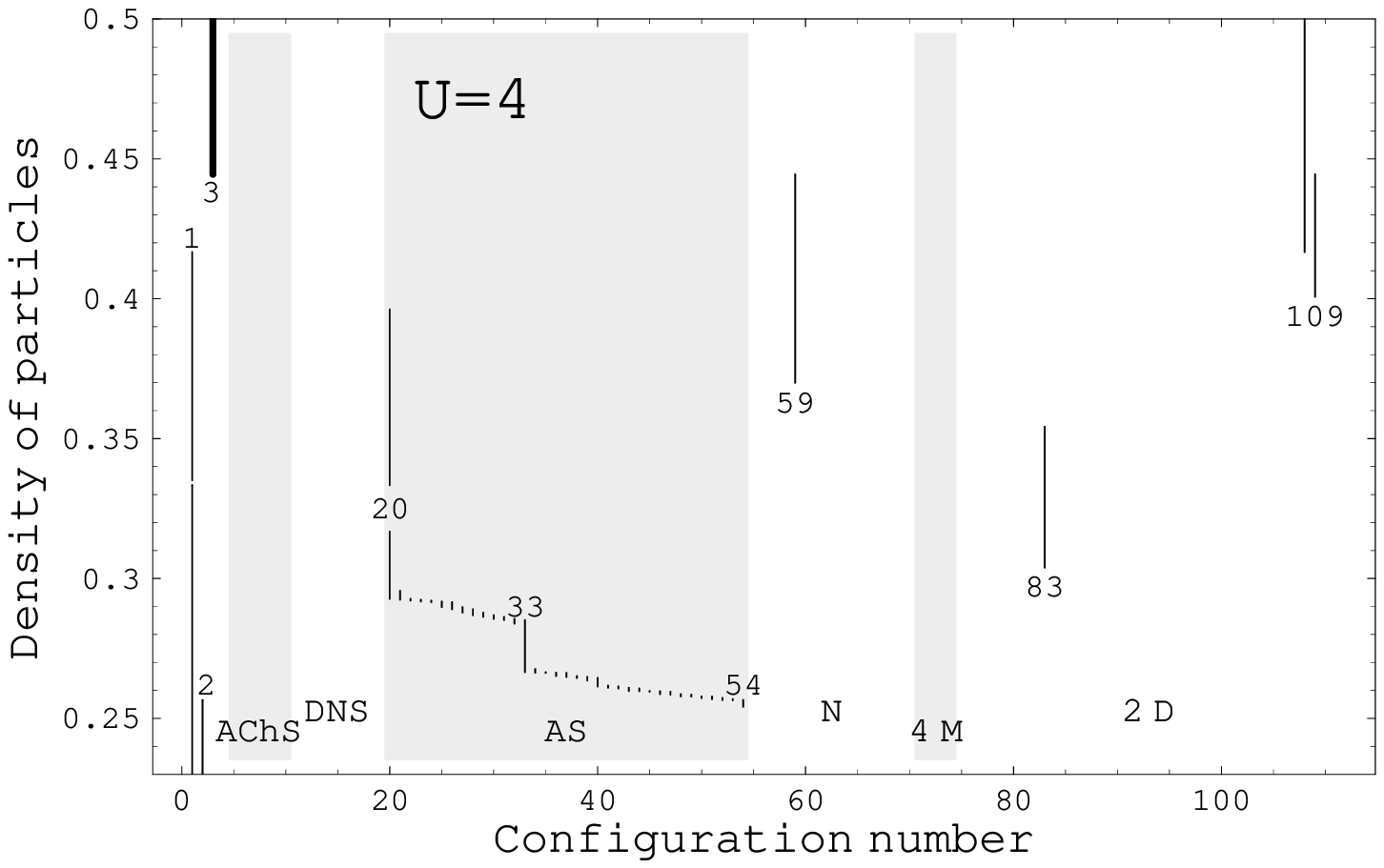}
\vskip 0.5in
Figure 6.

\newpage

\epsfxsize=4.4in
\epsffile{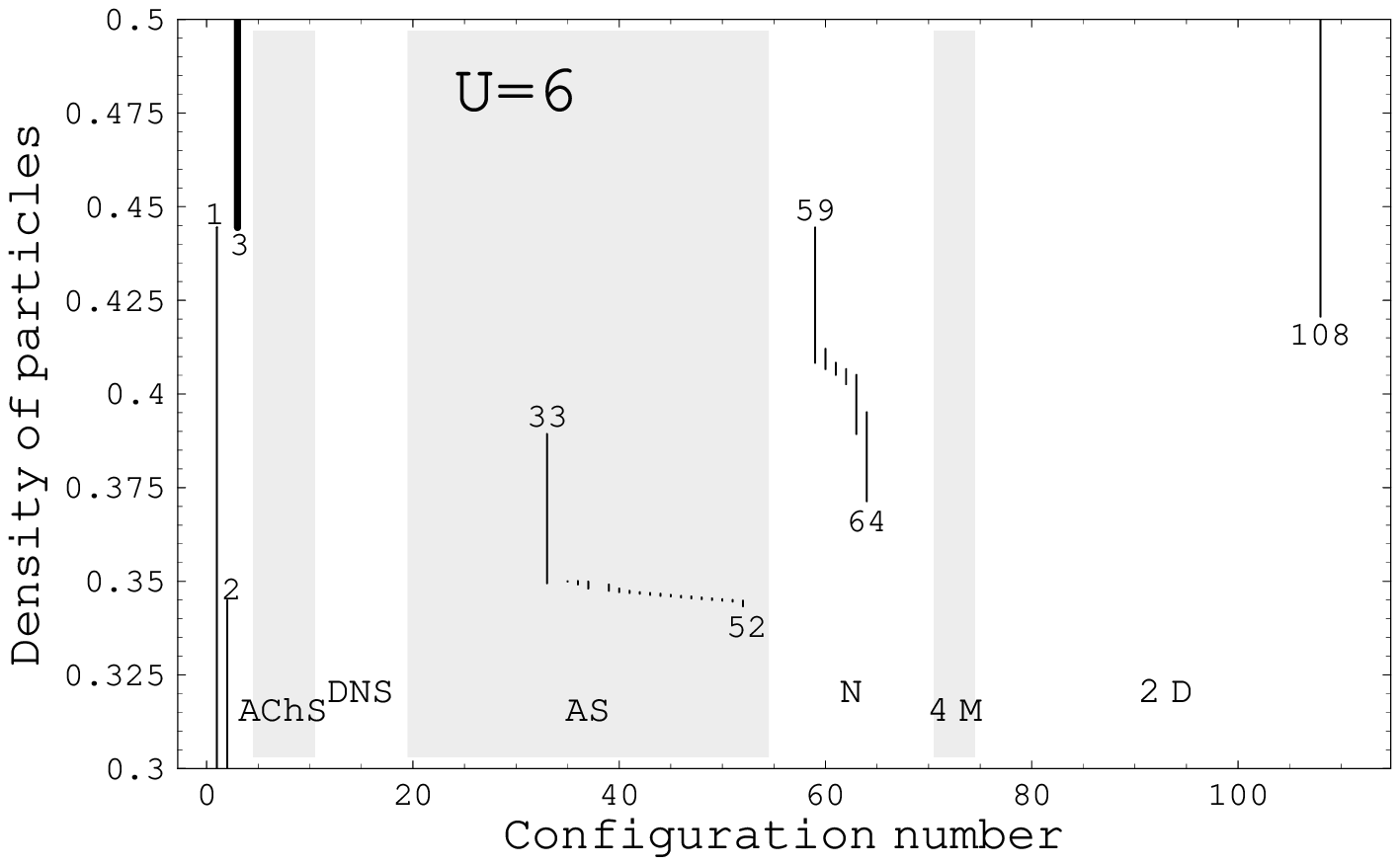}
\vskip 0.5in
Figure 7.

\newpage

\epsfxsize=4.4in
\epsffile{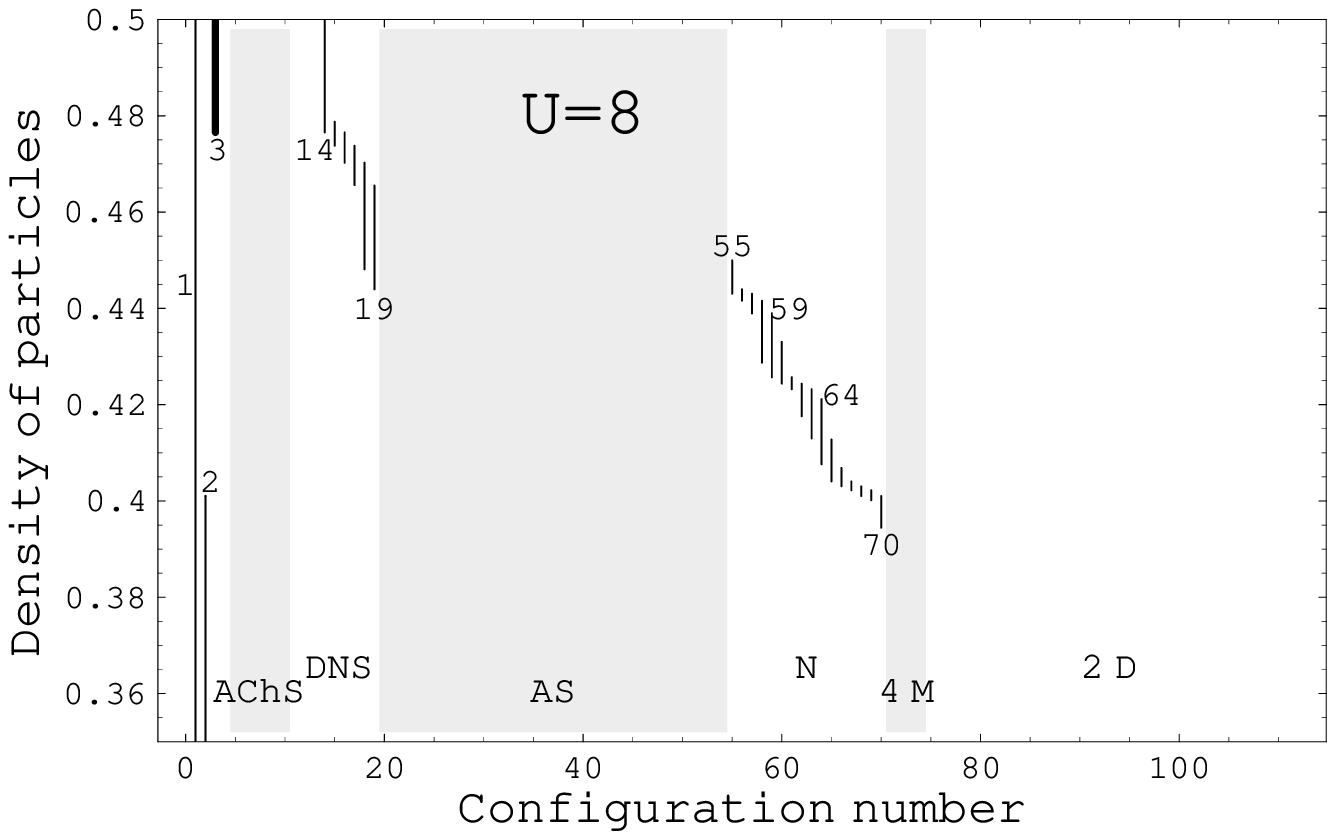}
\vskip 0.5in
Figure 8.

\newpage

\epsfxsize=4.4in
\epsffile{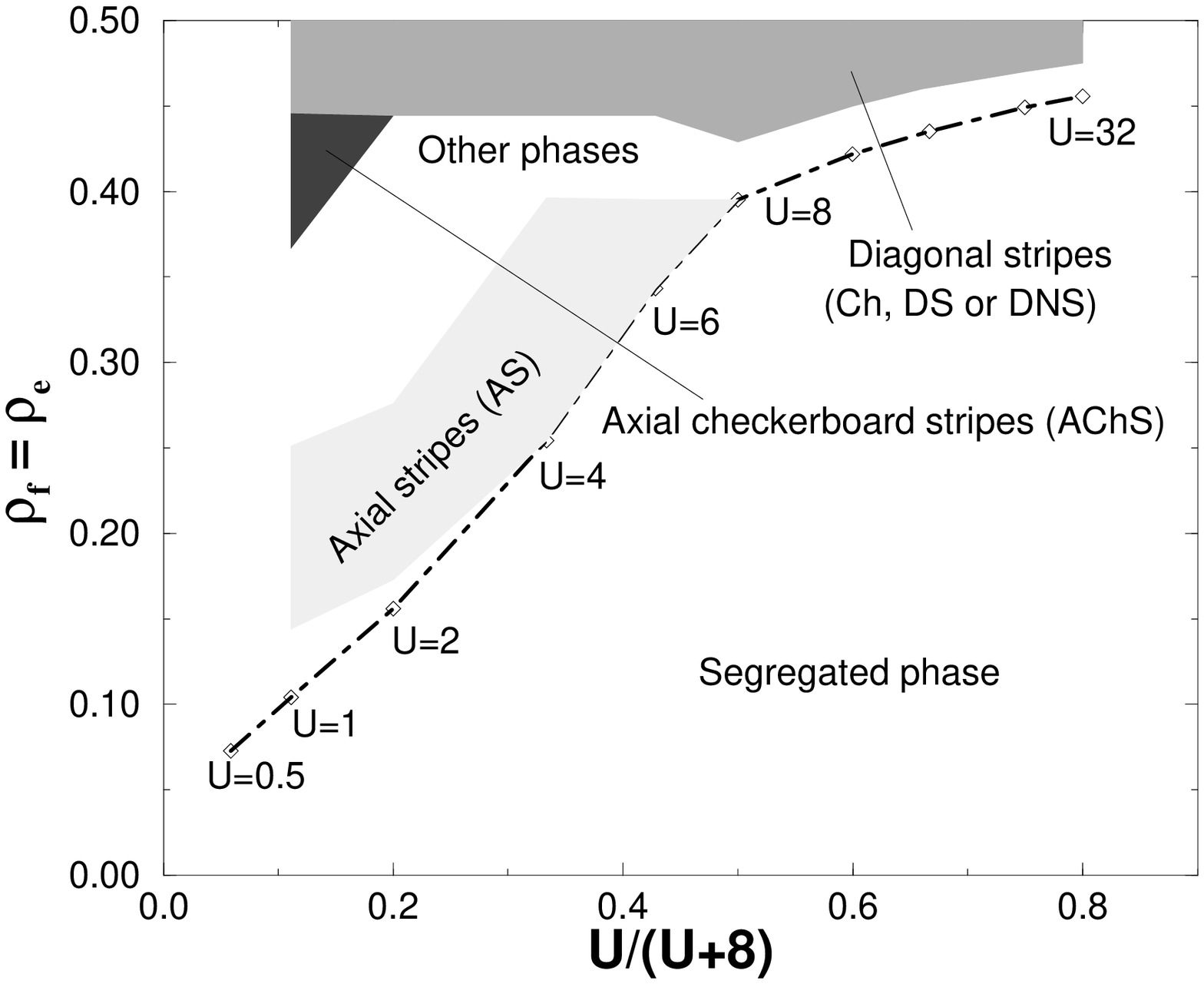}
\vskip 0.5in
Figure 9.

\end{document}